# Awareness of Technological Isomorphism: Integrating AI into Elementary Mathematics Teaching on Data and Prediction,A Case Study of the Compound Line Graph

**Li Li[1], Yu Cao[2]**

[1]Hefei No. 62 Middle School, Hefei, Anhui, China
[2]Anhui NARI ZT Electric Co., Ltd., Hefei, Anhui, China

**Abstract**

The deep integration of Artificial Intelligence (AI) into elementary mathematics education necessitates a conceptual tool capable of explaining students' cognitive transition from disciplinary knowledge to AI understanding. This study proposes a novel core concept, "Awareness of Technological Isomorphism, " defined as a student's metacognitive realization that their own mathematical cognitive operations (e.g., observing trends, inducing patterns, and making predictions) share an underlying logical structure with AI technical operations (e.g., pattern recognition and predictive modeling). This awareness, in turn, facilitates cognitive transfer from disciplinary mathematics to AI comprehension. Underpinned by transfer learning and metacognitive theories, this study clarifies the distinct essence of this concept from traditional "computational thinking." We demonstrate the explanatory power of this framework in two ways: elucidating the mechanism of students' cognitive leap from mathematics to AI, and guiding instructors to identify "isomorphic interfaces" within disciplinary curricula. On this basis, a three-stage pedagogical pathway--spanning "Perception, Comprehension, and Creation"--is constructed alongside a corresponding evaluation rubric. This framework is empirically validated through a case study based on the "Compound Line Graph" lesson from a fifth-grade mathematics textbook in China, offering a highly replicable operational framework for the deep convergence of disciplinary instruction and AI literacy education.

## 1.Introduction

In 2024, the General Office of the Ministry of Education of China issued the Notice on Strengthening Artificial Intelligence Education in Primary and Secondary Schools, mandating the basic universalization of AI education at the K-12 stage by 2030 [1]. Subsequently, provincial and municipal authorities, including Beijing and Shandong, released local implementation plans to integrate AI education into the comprehensive student evaluation system. Guided by these initiatives, a top-down transformation of AI literacy education is currently accelerating across primary and secondary schools nationwide.

However, a closer examination of elementary mathematics classrooms reveals a less optimistic reality. Currently, the integration of AI into mathematics instruction faces two prominent challenges. First, AI remains a mere "add-on" to the discipline, fostering a paradigm of "using AI" rather than "understanding AI." Teachers predominantly use AI as an efficiency tool for lesson planning or courseware generation. Consequently, students remain passive users of technology without understanding the underlying mechanisms of how AI achieves its outcomes—a predicament much like using a calculator without understanding the underlying arithmetic principles. Second, pedagogical pathways are highly fragmented and lack vertical articulation. AI education is traditionally confined to independent information technology courses, failing to connect with statistical knowledge in mathematics or experimental data analysis in science [2]. As a result, it is difficult for students to develop an intuitive, context-bound understanding of AI principles within core disciplinary environments.

The domain of "Statistics and Probability" in elementary mathematics provides a natural bridge to connect the two areas. From bar graphs in the third grade, to line graphs in the fifth grade, and ultimately to pie charts in the sixth grade, the continuum of mathematical thinking—characterized by "extracting trends and making judgments from data"—shares important similarities with the technical logic of machine learning, which focuses on "discovering patterns and making predictions from historical data." These two realms share a high degree of structural isomorphism[3]. If teachers can build on this shared structure to introduce AI predictive principles based on students' pre-existing mathematical knowledge, they can achieve the dual goal of helping students learn mathematics and understand AI concurrently.

However, to translate this isomorphic relation into actual teaching practice, a critical conceptual tool is still missing: What psychological mechanism enables students to transition from the mathematical operation of "observing trends and making judgments" to the technical understanding of "AI executing the same task"? Although previous studies have noted similarities between mathematical reasoning and AI thinking, limited attention has been paid to conceptualizing and operationalizing the specific cognitive mechanism that links the two [4]. Consequently, instructional designs frequently lack an explicit cognitive anchor, leaving educators unable to deliberately activate this cognitive channel during instruction. To address this theoretical gap, this study introduces the core concept of "**Awareness of Technological Isomorphism**."

Premised on this framework and using the primary mathematics continuum of "Data and Prediction" as the carrier, this study focuses on two pivotal research questions:

- RQ1: What is the theoretical connotation of "Awareness of Technological Isomorphism, " and how does it conceptually explicate students' cognitive leap from mathematical cognition to AI comprehension?
- RQ2: How can we utilize "Awareness of Technological Isomorphism" as an analytical framework to construct an "AI + Mathematics" pedagogical pathway aligned with students' cognitive development, and validate its efficacy through a concrete classroom case study?

## 2. Awareness of Technological Isomorphism: Conceptual Framework and Theoretical Foundations

### *2.1 Conceptual Definition*

The "Awareness of Technological Isomorphism" refers to a student's recognition during disciplinary learning that their own cognitive operations (e.g., observing data trends, inducing regularities, and making predictions) share an underlying logical structure with the technical operations of artificial intelligence (e.g., pattern recognition in training data and model-based trend prediction). Guided by this awareness, students can achieve a cognitive transfer from disciplinary knowledge to AI understanding [5].

This concept consists of three progressive levels. The first level is the "**existence of isomorphism**": trend-based reasoning in mathematics and trend prediction in AI inherently share a common underlying structure, which is "extracting rules from data to infer the unknown" [6]. The second level is the "**recognition of isomorphism**": under instructional guidance, students become aware that AI is performing a process based on the same underlying logic as the one they have just used. The third level is the "**internalization of isomorphism**": students internalize this recognition as a stable cognitive framework, forming a rational understanding that "AI principles are not mysterious—AI analyzes regularities in data just like I do." Consequently, they can actively apply this framework to comprehend other AI functions in new disciplinary contexts [5].

The notion of awareness in this study draws on epistemological discussions of the transition from unreflective engagement to conscious recognition [7]. When students observe trends and make judgments in a traditional mathematics class, they typically engage in these cognitive operations without consciously reflecting on them. However, when the teacher reveals that "AI is executing the same task, " students begin to recognize the isomorphic relationship between their own actions and the operations of AI. This represents a cognitive shift from unreflective cognition to conscious recognition [7]. It is precisely this transition that allows AI principles to naturally integrate into students' pre-existing cognitive structures as an intuitive extension, rather than being imposed as unfamiliar, external knowledge.

### *2.2 Theoretical Foundations and Conceptual Distinctions*

The "Awareness of Technological Isomorphism" is theoretically anchored in established learning and cognitive theories, while maintaining essential conceptual boundaries that distinguish it from existing frameworks.

#### 2.2.1 Transfer of Learning Theory.

Ausubel's cognitive structure transfer theory posits that meaningful learning occurs when new knowledge establishes non-arbitrary, substantive relations with an individual's pre-existing cognitive structures [8]. In the instructional context of "Data and Prediction, " students' pre-existing cognitive structures regarding trend estimation function as an "advance organizer" [9]. For instance, once students master the mathematical estimation that "if the temperature drops about 6 degrees every 30 minutes, it will reach approximately 72 degrees at 120 minutes, " they establish a baseline logic. When the instructor subsequently introduces how "AI predicts the fifth data point based on the previous four, " the predictive logic of AI does not enter the cognitive system as entirely unfamiliar, external knowledge. Instead, it is assimilated into the pre-existing structure of trend estimation [8]. The Awareness of Technological Isomorphism provides a theoretical explanation for why such transfer can occur, demonstrating that mathematical trend estimation and technological trend prediction share a similar underlying cognitive logic.

#### 2.2.2 Metacognitive Theory.

The concept of Awareness of Technological Isomorphism draws on metacognitive theory, as conceptualized by Flavell, which emphasizes that a learner's awareness and monitoring of their own cognitive processes are central to advanced cognitive development [5]. Awareness of Technological Isomorphism involves learners' recognition of their own cognitive processes and their specific relationship to AI operations. Here, students not only perform the skill of trend observation but also attain an explicit recognition that they are observing a trend, which subsequently develops into the realization that they and the AI are operating according to a similar underlying logic [5]. This three-tiered progression—moving from a raw cognitive operation to the recognition of structural isomorphism—mirrors the classic developmental trajectory of metacognition [10]. When a student spontaneously exclaims, "AI does not know the lid was opened, " it is precisely this metacognitive monitoring that empowers the awareness of isomorphism to transition from merely identifying structural alignment to critically discerning its operational boundaries.

2.2.3 Distinction from "Computational Thinking."

To validate the unique scholarly value of the proposed concept, it is crucial to establish its fundamental distinctions from the widely discussed construct of "Computational Thinking" (CT). As conceptualized by Wing, the primary orientation of CT is to enable students to "think like a computer scientist" through problem deconstruction, pattern recognition, abstraction, and algorithmic design [11]. The cognitive vector of CT moves from human cognition toward machine logic, aligning human thought processes with the operational mechanics of computers. In contrast, the Awareness of Technological Isomorphism focuses on helping learners recognize the similarities between their own reasoning processes and AI operations. Students are not required to learn to "think like a machine"; rather, they discover that their inherent cognitive operations are fundamentally mirrored by the machine's processes. The two constructs are complementary rather than mutually exclusive: computational thinking cultivates the ability to formulate and solve problems computationally, whereas the Awareness of Technological Isomorphism fosters an understanding of how AI relates to human cognitive processes. At the elementary education level, the latter may be more accessible to elementary school students given their stage of cognitive development.

*2.3 The Explanatory Power of the Concept*

The concept of Awareness of Technological Isomorphism offers a useful lens for understanding three important issues in AI-integrated mathematics education:

**First, it explicates the psychological mechanism underlying students' cognitive transition from disciplinary knowledge to AI understanding.** This concept shifts the locus of recognition regarding isomorphic relations from the researcher to the learner. Rather than being a ready-made conclusion delivered by the instructor, the isomorphic relationship becomes something learners come to recognize through their own cognitive activity. This shift effectively transforms "isomorphism" from an analytical category utilized merely by researchers into a dynamic, experiential cognitive event for the students themselves [5].

**Second, it provides an operational analytical framework for instructional design**. By consciously searching for points of isomorphism between disciplinary cognition and AI operations, educators can evaluate mathematics textbooks with heightened intentionality during lesson planning. They can systematically analyze which specific cognitive operations performed by students structurally align with which technical operations of AI, and determine the precise pedagogical stage and strategy needed to activate students' awareness of technological isomorphism. This analytical framework enables teachers to move away from a trial-and-error approach to AI integration, transitioning instead toward theoretically guided instructional design [8].

**Third, it addresses the limited explanatory power of "Computational Thinking" within the context of elementary education.** Many elementary school students are still developing the capacity for formal operational reasoning, which can create a substantial cognitive barrier to the traditional CT mandate of "thinking like a computer scientist." The Awareness of Technological Isomorphism provides a more accessible route: it does not demand a fundamental alteration of students' existing modes of thought, but rather requires them to recognize that their inherent reasoning processes are already structurally continuous with AI

logic [5]. This perspective offers a developmentally appropriate pathway for AI literacy education in elementary school settings.

## 3. Instructional Pathway Design: The Three-Stage Model of "Perception, Comprehension, and Creation"

### *3.1 The Logical Starting Point: Points of Isomorphism within the Mathematical "Data and Prediction" Continuum*

Within the domain of "Statistics and Probability" in elementary mathematics, there exists a well-defined curriculum continuum centered on "Data and Prediction": progressing from learning bar graphs in the third grade, to simple line graphs in the fourth grade, to compound line graphs in the fifth grade, and ultimately to pie charts and integrated applications in the sixth grade. The core cognitive requirement across this continuum is to extract trends and make informed judgments based on data. This mathematical reasoning process shares a structurally isomorphic logic with machine learning operations, which focus on discovering patterns and making predictions from historical data (see Table 1).

**Table 1.** Isomorphic Relationships Between Mathematical Reasoning and AI Operations.

| Mathematical Cognitive Operations | AI Technical Operations | Points of Isomorphism |
|---|---|---|
| Observing trends and variations in data | Pattern recognition within training data | Extracting underlying rules from data |
| Making estimations/predictions based on trends | Generating predictive values based on models | Utilizing extracted rules to extrapolate the unknown |
| Evaluating the reliability of mathematical estimations | Evaluating model prediction accuracy | Recognizing the inherent uncertainty of predictive outcomes |

The third dimension in Table 1 is particularly critical. When students perform mathematical estimations based on observed trends, terms such as "approximately" or "roughly" reflect an implicit understanding of uncertainty. Similarly, AI predictions are also subject to uncertainty, and their reliability may vary depending on available data and contextual conditions.

This isomorphic relationship serves as both the logical starting point for our instructional pathway and the cognitive foundation for the Awareness of Technological Isomorphism. The pre-existing ability of students to observe trends and make judgments in a traditional mathematics classroom provides an important cognitive foundation for understanding AI predictive principles. Consequently, educators do not need to introduce AI concepts through an entirely isolated curriculum. Instead, they can build on students' established mathematical knowledge to construct a cognitive bridge that connects human empirical judgment with machine-based data computation. The concept of Awareness of Technological Isomorphism provides the basis for constructing this cognitive bridge.

### *3.2 The Three-Stage Instructional Pathway Design*

Drawing on the "Black Box, Grey Box, White Box" cognitive progression model proposed in the Curriculum Scheme and Implementation Recommendations for AI Education in Primary and Secondary Schools by Beijing Normal University (2025) [12], and integrating it with the conceptual architecture of mathematics curricula, this study constructs a three-stage pedagogical pathway: "Perceiving, Comprehending, and Creating." Grounded in the developmental logic of the Awareness of Technological Isomorphism, the three stages are explicated as follows:

**Stage 1: Perceiving AI (The Black Box Stage, Grades 1–3).** In this introductory phase, students engage with technology primarily through application, utilizing AI-powered graphing tools to complete data visualization tasks. The experiential focus centers on how "AI accelerates graph generation, " while the instructional core remains primarily anchored in foundational mathematical and statistical knowledge. This stage serves as a preparatory phase, during which students accumulate relevant cognitive experiences through empirical mathematical activities of extracting trends from data, thereby laying the experiential foundation for the later development of isomorphic awareness.

**Stage 2: Comprehending AI (The Grey Box Stage, Grades 4–5).** The core pedagogical mechanisms governing this stage are "**dual-track comparative analysis**" and "**reflection on prediction failures**." Students first complete a manual mathematical investigation following the sequence of "observing trends analyzing regularities making judgments." They then compare this manual experience with the prediction process of AI, addressing how the machine reaches a prediction through a similar underlying logic. Subsequently, through the introduction of scenarios where AI predictions fail, students critically engage with the limitations of AI systems. This stage represents the critical "**activation phase**." The dual-track comparison activates the initial perception of isomorphism, while the reflection on prediction failures deepens this insight, serving as the pivotal turning point for the students' cognitive leap.

**Stage 3: Creating AI (The White Box Stage, Grades 5–6).** In this advanced phase, students actively experience the entire machine learning pipeline—"providing datasets enabling machine learning verifying predictive accuracy"—using graphical programming tools. This hands-on experience allows them to recognize the similarities between the logic of machine learning and their own processes of learning from data. This stage serves as the "**internalization phase**, " where the realization that "AI learns from data in ways that resemble my own reasoning processes" shifts from a newly activated awareness into a stable cognitive framework, solidifying as the students' default perspective when encountering AI technologies. It should be noted that the empirical case validation in this study focuses specifically on Stage 2.

The overall structure of the pathway and the cognitive progression of this three-stage model are presented in Figure 1.

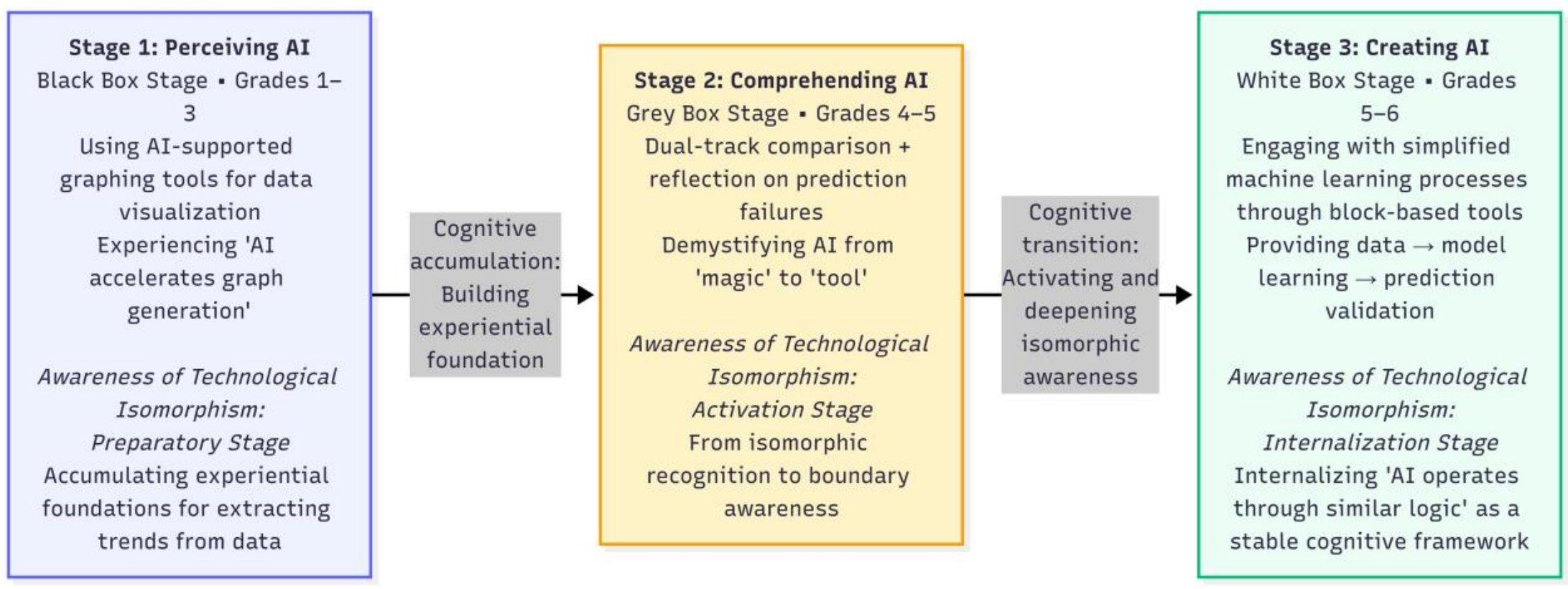

**Figure 1.** The Three-Stage Instructional Pathway of "Perceiving–Comprehending–Creating " and the Development of Awareness of Technological Isomorphism

The implementation of the aforementioned instructional pathway adheres to two guiding principles:

- **Mathematical Primacy (Mathematics-Centered Orientation):** The integration of AI is designed to serve the deepening of mathematical concepts. The activation of the Awareness of Technological Isomorphism is built upon students' thorough and robust mastery of mathematical reasoning.
- **Cognitive Progression:** The pathway follows a deliberate trajectory moving from "using AI, " to "understanding AI, " and ultimately to "engaging with AI creation." This progression maps systematically onto the complete developmental arc of isomorphic awareness, transitioning from a preparatory phase, through an activation phase, to an internalization phase.

## 4. Instructional Case Study: "Compound Line Graphs" (Jiangsu Education Edition, Grade 5)

Focusing on Stage 2 ("Comprehending AI"), this section explicates the classroom implementation of the three pedagogical steps: mathematical empirical investigation, AI prediction comparison, and reflection on prediction failures. Through an analysis of classroom dialogue segments, this section illustrates the specific mechanisms through which the Awareness of Technological Isomorphism is activated and deepened.

*4.1 Mathematical Justification for Case Selection*

Selecting the unit "Compound Line Graphs" from the Grade 5 mathematics textbook (Jiangsu Education Edition) as the core case for Stage 2 is predicated not on its utility for technological demonstration, but on the following rigorous mathematical considerations:

4.1.1 The Core Mathematical Value Lies in Comparison and Prediction.

The textbook introduces an experimental dataset tracking water temperature inside two different vacuum flasks (see Table 2). The instructional goal requires students to achieve a cognitive transition from raw data reading, to trend analysis, and ultimately to comparative judgment. Specifically, the task of determining "approximately how many minutes it takes for the water temperature to drop to 70°C" constitutes a problem of **interpolation and estimation**. It requires students to infer an unknown data point from an established trend—an operation that shares a structurally isomorphic logic with AI prediction.

**Table 2.** Experimental Data of Temperature Changes in Two Types of Vacuum Flasks.

| Elapsed Time (min) | 0 | 30 | 60 | 90 | 120 | 150 |
|---|---|---|---|---|---|---|
| Stainless Steel Flask Temp (°C) | 95 | 90 | 84 | 78 | 72 | 68 |
| Ceramic Flask Temp (°C) | 95 | 74 | 58 | 50 | 45 | 41 |

4.1.2 The Mathematical Essence of "Approximately" is Trend-Based Estimation.

By observing the compound line graph, students can discern that the water temperature in the stainless steel flask decreases by approximately 6°C every 30 minutes. Based on this regularity, they estimate that the 70°C mark occurs around the 130-minute point. This mathematical reasoning of estimating based on trends shares the same underlying logical structure as AI predictive models, which infer subsequent values from patterns detected in historical datasets.

4.1.3 Methodological Grounding of the Instructional Design.

During the instructional design process, an exploratory comparison was conducted between students' manual estimation strategies and AI-generated predictions based on the textbook dataset. Upon observing that the algorithm's predictive values closely aligned with students' manual estimations, it became evident that a student plotting points on a graph to estimate an unobserved value and a computer processing historical data to output a model prediction are engaged in operations that share a similar underlying logical structure. This insight reshaped the approach to technology integration: rather than introducing AI through an isolated, extraneous curriculum, educators can enable students to recognize AI operations within the mathematical activities they have already mastered. Consequently, trend-based estimation serves as the cognitive bridge connecting human reasoning with machine computation—a bridge theoretically conceptualized in this study as the Awareness of Technological Isomorphism.

*4.2 Design of Instructional Objectives*

4.2.1 Mathematical Literacy Objectives (Core)

- Students will be able to identify and compare the upward and downward trends represented by two datasets in a compound line graph.
- Students will be able to make reasonable estimations based on observed trends, thereby strengthening their data awareness and reasoning abilities.

4.2.2 AI Cognitive Objectives (Extended)

- Students will be able to explain, in their own words, the basic logic underlying AI trend prediction, recognizing that AI identifies patterns in data in much the same way that students estimate trends from graphs, and that the two processes share a similar underlying logical structure.
- Students will be able to identify situations in which AI predictions may fail, thereby developing an informed and critical understanding of the capabilities and limitations of AI tools.

*4.3 Instructional Procedures*

The lesson was organized into three instructional segments: Mathematical Inquiry (15 minutes), AI Prediction Comparison (8 minutes), and Reflection on Prediction Failures (5 minutes).

The underlying pedagogical logic and the cognitive progression across these three segments are illustrated in Figure 2.

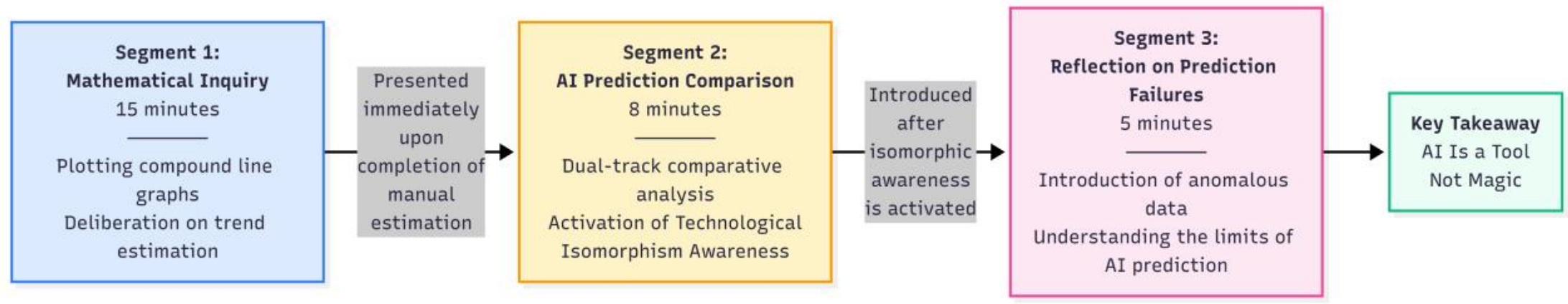

**Figure 2.** Instructional Sequence and Cognitive Progression Across Three Segments.

*4.4 Classroom Implementation and Discourse Analysis*

4.4.1 Segment 1: Mathematical Inquiry (15 Minutes)

Based on the experimental data provided in the textbook, students independently constructed the compound line graphs and subsequently engaged in group discussions surrounding three progressive prompts:

- Prompt 1 (Data Extraction): What were the temperature differences between the two flasks at the 60-minute and 120-minute marks, respectively?
- Prompt 2 (Trend Estimation - The Core Component): Approximately how many minutes did it take for the water temperature in the stainless steel flask to drop to 70°C? What about the ceramic flask?
- Prompt 3 (Comparative Judgment and Extended Discovery): Which flask exhibits superior thermal insulation performance? What other patterns can you discern from the graph?

During this segment, the teacher guided students through the complete cognitive trajectory of "observing trends -> identifying regularities -> formulating estimations." The instructional focus was intentionally placed on the mathematical essence of the word "approximately" in Prompt 2, leading students to realize that estimation is not arbitrary guessing, but rather a reasoned inference grounded firmly in mathematical trends. This mastery established the vital cognitive foundation for activating the Technological Isomorphism Awareness; students must first internalize the ability to analyze visual trends before they can recognize that AI systems execute a comparable operation.

4.4.2 Segment 2: AI Prediction Comparison (8 Minutes)

To initiate this segment, the teacher posed an inquiry: "If we only provide the AI system with data from the first four time points, can it infer the water temperature at the fifth time point?"

The teacher then executed a demonstration program, displaying two overlapping trajectories on the screen: one representing the actual values plotted manually by the students, and the other representing the predictive curve generated by the AI based solely on the historical data of the first four intervals.

The following is a selected transcript segment from the classroom discourse:

**Teacher:** The AI predicted a temperature of 72.3 ° C, while the actual recorded temperature was 72°C. What do you think of its prediction?

**Student 1 (S1):** It's so accurate! The difference is only 0.3 degrees!

**Teacher:** What was your manual estimation just now?

**S1:** I estimated it to be somewhere between 72°C and 73°C.

**Teacher:** And what was the basis for your estimation?

**S1:** I noticed that the temperature drops by about 6°C every 30 minutes. At 90 minutes, it was 78°C, so subtracting another 6°C gives exactly 72°C.

**Teacher:** (Turning toward the entire class) Then what do you think the AI relied on to make its guess? It was also calculating "how much the temperature drops every 30 minutes, " except it utilized a more precise mathematical formula. Ultimately, you and the AI are relying on a similar way of identifying trends in data.

(Several students nod in agreement; one student commented quietly: "So it's kind of doing what I did.")

**Teacher:** Let me ask you once more—why was the AI able to predict so accurately?

**Student 4 (S4):** Because it is just like us; it is also looking for patterns hidden within the data.

The significance of this interaction lies in the fact that the teacher did not directly provide the explanation. Instead, students first articulated the basis of their own estimations before recognizing that AI prediction relies on a similar underlying logic. When S4 stated that "it is also looking for patterns hidden within the data, " a cognitive shift occurred from "I am estimating" to "AI is also estimating."

**Indicators of Emerging Technological Isomorphism Awareness**

The vocalizations—"So it's kind of doing what I did" and "Because it is just like us"—can be regarded as observable indicators that Technological Isomorphism Awareness had begun to emerge. While students intensively practiced trend estimation during Segment 1, their awareness remained in a **tacit state** (unreflected and spontaneous). Segment 2 utilized a dual-track comparative analysis to bridge the two domains; by first witnessing the convergence of outcomes and subsequently tracing it back to the equivalence of processes, students externalized their realization through the phrase "it is just like us." This verbalization may be interpreted as an observable marker of the transition from a preparatory stage to an activation stage.

The classroom observation suggests that two conditions may have contributed to this activation:

- **Internalized Mastery:** Students appeared to have internalized the mathematical operation of trend estimation, providing a robust cognitive anchor for subsequent reflection.
- **Pedagogical Scaffolding:** The teacher refrained from providing a direct explanation, creating opportunities for students to discover the structural correspondence independently. The former guarantees the cognitive foundation, while the latter ensures that the realization is an autonomous awakening ("awareness") rather than direct instruction.

4.4.3 Segment 3: Reflection on Prediction Failures (5 Minutes)

The teacher introduced a new secondary dataset (see Table 3), which documented a contextual anomaly: the lid of the vacuum flask was deliberately opened at the 90-minute mark of the experiment.

**Table 3.** Experimental Data Incorporating a Contextual Anomaly.

| Elapsed Time (min) | 0 | 30 | 60 | 90 | 120 | 150 |
|---|---|---|---|---|---|---|
| Water Temperature (°C) | 95 | 88 | 81 | 74 | 55 | 40 |

The teacher then posed a question: "If the AI only analyzed the first four data points, what temperature would it predict for the fifth point?" Based on the previous trend, students initially estimated it to be approximately 67°C. However, the actual recorded temperature was 55°C, as the opening of the lid introduced an abrupt environmental shift.

The following is a selected transcript segment from the classroom discourse:

**Teacher:** The AI predicted 67°C, but the actual temperature was 55°C! The prediction error increased significantly by 12°C! Why did the model produce such a large deviation here?

**Student 2 (S2):** Because it didn't know the lid was opened!

(The classroom erupts into laughter)

**Teacher:** Exactly. The AI only saw the numbers from the first four intervals. Did it have any way of knowing someone opened the lid?

**S2:** No, it didn't!

**Teacher:** Then who knew the lid was opened?

**Student 3 (S3):** We did!

**Teacher:** So, under what circumstances might our own guesses complement the AI's predictions?

**Student 5 (S5):** When we know real-world facts that the AI doesn't. For example, we know the lid was opened, but the AI cannot see that.

**Student 6 (S6):** Also, if the data suddenly shifts, the AI won't know how to adjust because it doesn't know what happened. But we can guess what went wrong in the real world.

**Teacher:** Well said. So, how should we view AI?

**Student 7 (S7):** AI is just a tool; it doesn't actually know everything.

**S2:** (Adding on) It only looks at the numbers.

The essence of this interaction appears to reflect S2's spontaneous exclamation: "Because it didn't know the lid was opened!" This vocalization may indicate that the students had begun moving past an intuitive, uncritical fascination with technological capability. They appeared to realize that while AI predictions depend strictly on historical datasets, human inference can complement model-based prediction when contextual information is missing from the dataset. The subsequent articulations by S5 and S6 suggest that students were utilizing the boundaries of structural correspondence to explain the operational limitations of AI predictive models.

**Indicators of Deepening Technological Isomorphism Awareness**

During Segment 2, while students recognized that "AI analyzes trends in a way similar to human estimation, " such a realization risks sliding into the misconception that technological predictions are inherently infallible. Classroom observations in Segment 3 suggest that S2's statement marked a pivotal shift wherein Technological Isomorphism Awareness transitioned from basic recognition to boundary awareness. Students demonstrated an understanding that the information boundaries for human reasoning and AI operations are fundamentally distinct: human inference has access to situational contexts (the opening of the lid) unavailable to the model, whereas the AI model remains confined strictly to numerical inputs.

This distinction illustrates a potential cognitive progression from merely recognizing structural similarity to understanding its operational prerequisites. This trajectory traces the plausible developmental movement of Technological Isomorphism Awareness within the Stage 2 curriculum: progressing from a preparatory stage (Segment 1), through an activation stage (Segment 2), and ultimately toward a deepening phase (Segment 3), as conceptualized in Figure 3.

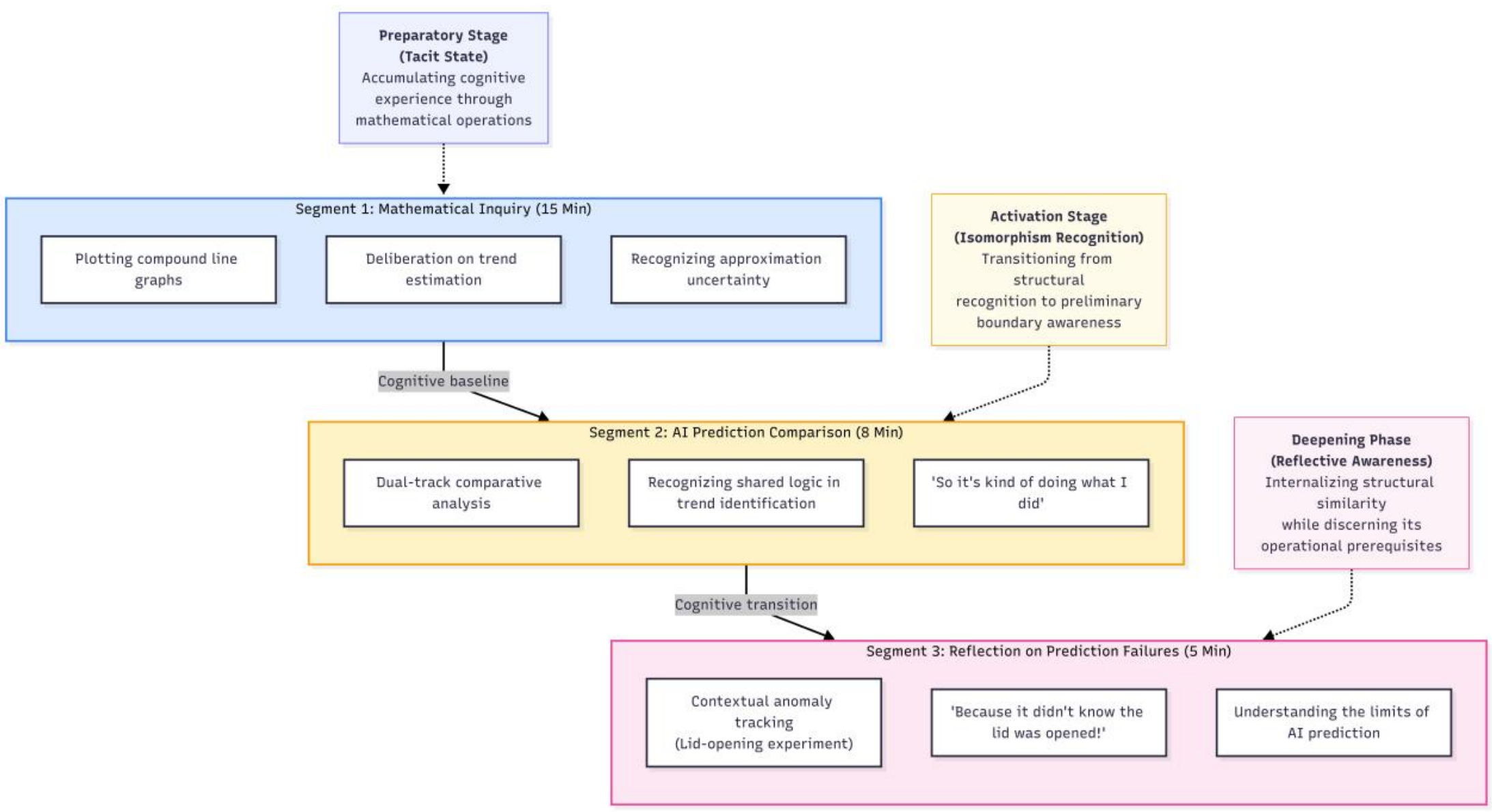

**Figure 3.** The Classroom Trajectory and Developmental Trajectory of Technological Isomorphism Awareness in Classroom Practice.

This instructional segment may illustrate a pedagogical shift from an uncritical reliance on AI toward a more informed and balanced perspective: recognizing that predictive models operate on the basis of historical data, while human reasoning may incorporate contextual information that is not explicitly represented in the dataset. This progression extends technological isomorphism awareness from structural recognition to boundary-sensitive understanding, highlighting the conditions under which similarity between human and AI reasoning is maintained or disrupted.

.

*4.5 Design of Instructional Evaluation*

The evaluation framework for this lesson focuses on three progressive dimensions, with each dimension structured across three proficiency levels (see Table 4). Within this rubric, mathematical literacy serves as the foundation, with AI cognition and ethical awareness treated as derived dimensions.

**Table 4.** Instructional Evaluation Rubric for Technological Isomorphism Awareness.

| Evaluation Dimension | Level 1 | Level 2 | Level 3 | Assessment Methods |
|---|---|---|---|---|
| Data Literacy & Statistical Reasoning | Students can extract and recall specific, isolated data points from the compound graph. | Students can formulate rational estimations based on trends and articulate the underlying mathematical basis. | Students can compare trend variances between two datasets and deduce evidence-based judgments. | Accuracy of manually plotted graphs; Classroom discourse analysis. |
| AI Cognitive Understanding | Students can state that the AI model generated a predictive value based on a given set of inputs. | Students can explain that AI predictions are derived from identifying regularities and patterns within data. | Students can articulate that algorithmic prediction and human trend estimation share a fundamentally similar underlying logic. | Post-test open-ended questionnaire. |
| Ethical and Boundary Awareness | Students can identify and note that the AI prediction deviated significantly from the actual value. | Students can explain that the predictive deviation was caused by the model's lack of access to the lid-opening context. | Students can synthesize that while AI relies strictly on numerical patterns, human inference can complement algorithmic predictions in volatile real-world contexts. | Verbal behavior during reflection on prediction failures. |

Level 3 of the AI Cognitive Understanding dimension is treated in this study as an observable indicator of the activation of Technological Isomorphism Awareness, while Level 3 of the Ethical and Boundary Awareness dimension indicates progression toward boundary-sensitive understanding. Consequently, this rubric functions as a structured tool for evaluating student performance within classroom practice.

*4.6 Classroom Observations and Pedagogical Outcomes*

Classroom observations indicate that the implementation of this instructional design yielded a three-tiered progressive outcome:

- **First,** during the dual-track comparative analysis segment, students spontaneously externalized a form of cognitive resonance, expressing that the AI was identifying trends in a manner comparable to their own reasoning. Verbalizations such as "So it's kind of doing what I did" and "Because it is just like us" were recorded, indicating initial activation of Technological Isomorphism Awareness.

- **Second,** during the reflection on prediction failures segment, students not only recognized that the algorithmic prediction deviated substantially from observed values but also accurately attributed this deviation to situational data external to the model. S2's exclamation ("Because it didn't know the lid was opened!") and S6's observation ("we can guess what went wrong in the real world") were prominent behavioral markers, highlighting perceived boundaries between human and model-based inference.
- **Third,** during the post-lesson data collection, several students reflected that AI technologies are fundamentally grounded in mathematical structures rather than being enigmatic entities, suggesting an early tendency toward generalization of this awareness into more comprehensive technological perceptions.

Educators can utilize the rubric in Table 4 to conduct qualitative assessments through structured classroom observations and brief post-lesson interviews. Particular pedagogical attention should be directed toward whether students spontaneously produce contextual attributions—such as "Because it didn't know the lid was opened!"—during anomalous data segments, as these verbal behaviors mark the empirical shift toward boundary awareness.

### *4.7 Recommendations for Instructional Implementation*

To facilitate the transferability and replication of this curriculum framework, practitioners are encouraged to consider the following implementation recommendations:

#### 4.7.1 Prerequisite Technical and Cognitive Conditions

- Technical Infrastructure: The classroom must be equipped with multimedia display capabilities, and instructors are encouraged to execute and debug the predictive demonstration software prior to the session.
- Cognitive Anchors: It is recommended that this lesson be introduced after students have mastered data extraction and trend analysis using single-line graphs, thereby ensuring a reliable baseline for interpreting compound line graphs.

#### 4.7.2 Pacing and Temporal Allocation

Within a standard 40-minute instructional period, the temporal allocation should be carefully managed to maintain the cognitive coherence of each phase:

- Mathematical Inquiry (15 minutes): This segment serves as an essential foundation of the lesson, ensuring students have sufficient time to internalize visual trend operations.
- AI Prediction Comparison (8 minutes): The AI prediction process should be introduced immediately upon the completion of manual estimations to maximize the cognitive dissonance and highlight the structural correspondence.
- Reflection on Prediction Failures (5 minutes): As a key component for conceptual deepening, this segment is designed to provide students with space for spontaneous deliberation, dialogue, and collaborative reflection on the boundary conditions of predictive models.

## 5. Discussion, Limitations, and Future Directions

### *5.1 Research Contributions*

The primary contribution of this study lies in the proposal and conceptualization of "Technological Isomorphism Awareness." At the phenomenological level, this construct reframes the alignment between mathematical reasoning and AI operations from an observational phenomenon into a potentially cultivable cognitive construct, thereby positioning structural isomorphism as a student-facing cognitive experience rather than a researcher-only descriptive category. At the mechanistic level, the framework offers a conceptual explanation of how students may transition from disciplinary knowledge to AI understanding, suggesting that this transition is mediated by metacognitive recognition of structural correspondences rather than direct instructional transmission. At the practical level, it introduces a design principle of "isomorphic interfaces, " encouraging educators to identify correspondences between mathematical operations and algorithmic behaviors and to strategically structure classroom activation points.

Utilizing this conceptual framework, this study achieved two empirical milestones:

- First, it established a three-stage developmental trajectory — progressing from a preparatory stage (tacit state), through an activation stage, to a deepening phase—thereby clarifying the distinct cognitive functions of each period in fostering isomorphism awareness.
- Second, it translated these theoretical tenets into an actionable, replicable lesson design centered on Compound Line Graphs, providing empirical, discourse-based observations that illustrate how this awareness is activated, calibrated, and consolidated in a real-world classroom setting.

Furthermore, this exploratory study opens new avenues for investigating how technological isomorphism awareness might be activated and nurtured across diverse disciplines (e.g., experimental reasoning in the natural sciences, or textual synthesis in language arts) and varied educational stages.

*5.2 Research Positioning and Limitations*

Methodologically, this inquiry is positioned as design-based research (DBR), with the central objective of developing theoretically driven instructional interventions and corresponding evaluation tools. Consequently, several limitations inherent to this qualitative case study approach must be acknowledged:

- First, regarding the three-stage framework, this study focused primarily on the fine-grained instructional design and empirical evaluation of the first two stages, leaving the final stage (the internalization and creative application phase) open for further operational development.
- Second, the evaluation of student cognitive shifts relied predominantly on qualitative discourse analysis and observational indicators. While these data sources provide deep contextual insights, future research should prioritize the development of standardized psychometric instruments to facilitate larger-scale, empirical validation.

*5.3 Teacher Competency Requirements for Curricular Scaling*

The broader scaling and transferability of the proposed three-stage trajectory depend significantly on three core teacher competencies:

- Disciplinary Insight: Instructors must possess the capacity to identify structural correspondences between foundational disciplinary knowledge and AI principles. The alignment between trend estimation and machine prediction illustrated in this study represents merely one such interface; similar correspondences exist within probability models, cluster analysis, and statistical distributions.

- Pedagogical Positioning: Instructors must strictly adhere to a discipline-centric approach, maintaining mathematical discipline-centricity while utilizing AI as a pedagogical scaffold. The activation of technological isomorphism awareness is strictly contingent upon the robust and thorough internalization of the underlying mathematical concepts.
- Technical Self-Efficacy: Instructors are required to proficiently operate state-of-the-art Large Language Models (LLMs) featuring automated data visualization capabilities, though advanced computer programming skills are not a prerequisite.

*5.4 Future Horizons*

Looking forward, research into technological isomorphism awareness can be meaningfully expanded in three directions:

- First, larger-scale classroom implementations and iterative optimization should be conducted across diverse regions and varying socio-economic school tiers to systematically examine the generalizability of this instructional framework, paying specific attention to how variations in students' mathematical baselines influence their cognitive progression during the activation and deepening phases.
- Second, future instructional designs should fully operationalize the third stage — "Creating AI"—by investigating the specific cognitive manifestations and scaffolding strategies that occur when students engage with block-based visual tools to experience the complete data pipeline: from collection and training to validation.
- Third, subsequent research must prioritize the development of standardized, validated instruments to complement qualitative classroom observations. Translating current qualitative indicators into quantifiable, comparable performance metrics will provide empirical methodology to support large-scale, cross-sectional, and longitudinal investigations.

A longer-term research direction is the development of a comprehensive AI-integrated instructional ecosystem for the domain of Statistics and Probability in primary education, with Technological Isomorphism Awareness serving as a foundational conceptual construct.